\documentclass[usenatbib]{mn2e}

\usepackage{amsmath}
\usepackage{ulem}
\usepackage{amssymb}
\usepackage{float}
\usepackage{graphicx}
\usepackage{lscape}
\usepackage{url}
\usepackage{natbib}
\usepackage{threeparttable}

\title[MHD Shock Reorientation]{Reorienting MHD Colliding Flows:\\
A Shock Physics Mechanism for Generating Filaments Normal to Magnetic Fields}

\author[E. Fogerty et al.]{
Erica Fogerty$^{1}$\thanks{E-mail:erica@pas.rochester.edu}, 
Jonathan Carroll-Nellenback$^{1}$, Adam Frank$^{1}$,
Fabian Heitsch$^{2}$, \newauthor
~Andy Pon$^{3}$ \\\\
$^{1}$ 206 Bausch \& Lomb Hall, Department of Physics \& Astronomy, University of Rochester, Rochester, New York, 14627, USA\\
$^{2}$ 3255 Phillips Hall, Department of Physics \& Astronomy, University of North Carolina, Chapel Hill, North Carolina, 27599, USA\\
$^3$ Department of Physics \& Astronomy, University of Western Ontario, London, Ontario, N6A 3K7, Canada }

\begin{document}

\date{Submitted 2016 September} 


\maketitle

\label{firstpage}

\begin{abstract}

We present numerical simulations of reorienting oblique shocks that form in the collision layer between magnetized colliding flows. Reorientation aligns post-shock filaments normal to the background magnetic field. We find that reorientation begins with pressure gradients between the collision region and the ambient medium. This drives a lateral expansion of post-shock gas, which reorients the growing filament from the outside-in (i.e. from the flow/ambient boundary, toward the colliding flows axis). The final structures of our simulations resemble polarization observations of filaments in Taurus and Serpens South, as well as the integral-shaped filament in Orion A. Given the ubiquity of colliding flows in the interstellar medium, shock reorientation may be relevant to the formation of filaments normal to magnetic fields.






\end{abstract}

\begin{keywords}
{\it magnetohydrodynamics\/}\verb"(MHD)" -- ISM: kinematics and dynamics -- ISM: structure -- ISM: clouds -- stars: formation
\end{keywords}

\section{Introduction}

Colliding gas flows and filaments are commonly found in star forming regions. Converging flows have been detected surrounding molecular gas in Taurus \citep{ballesteros1999}, the Sh 156 and NGC 7538 molecular clouds \citep{brunt2003}, and star forming filaments in Serpens South \citep{kirk2013, fernandez2014}. On the largest scales, they can arise from supernovae, energetic winds surrounding young stars and clusters, and the motion of galactic spiral arms through the intragalactic medium. 
On smaller scales, converging flows can take the form of accretion flows. Similarly, filaments are ubiquitous in star forming regions. Many have been found to contain young protostellar cores \citep{andre2010,arzoumanian2011,polychroni2012, lee2014}, and thus, are considered some of the earliest structures of star formation. 

\begin{figure}
    \includegraphics[width=\columnwidth]{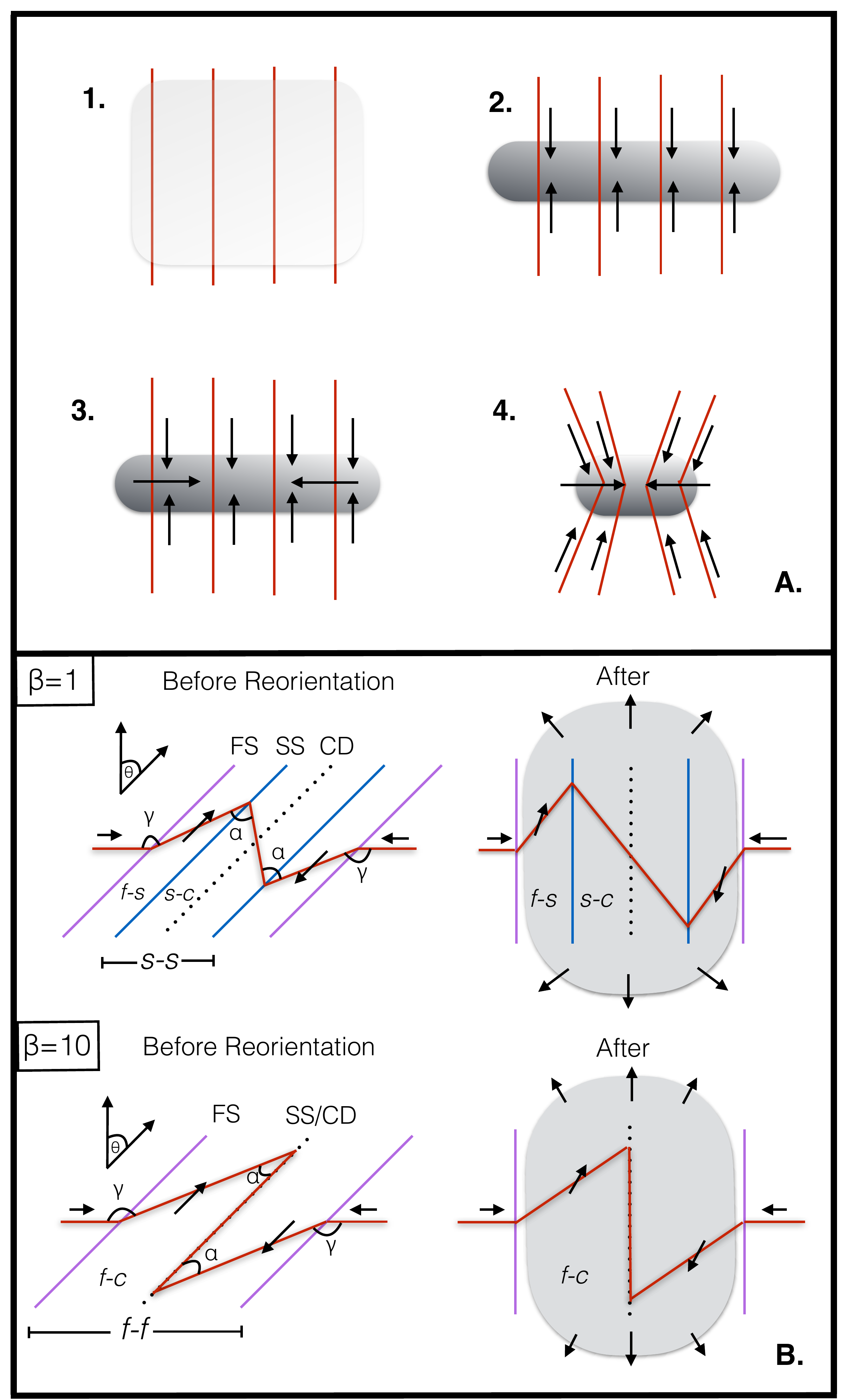}
    \caption{Mechanisms for generating filaments normal to magnetic field lines. \textit{Top Panel.}  Magneto-gravitational collapse. Jeans unstable gas will initially collapse along field lines (1-2). Once enough mass accumulates in the resultant filament, mass can flow along the filament and drag magnetic field lines inward (3-4).  \textit{Bottom Panel.} Reorientation of magnetized oblique shocks. Uppercase letters denote wave modes, and lowercase letters give regions (see text). Flow direction is given by the black arrows, and a representative magnetic field line is shown in red. The left and right panels give the initial and reoriented structures, respectively. The top and bottom rows give the moderate ($\beta=1$) and weak field ($\beta=10$) cases of this paper. The grey shaded region represents the growing filament.}
    \label{schematic_collapse}
\end{figure}

Observations of filaments indicate that they might be tied to colliding flows. Measurements of velocity gradients perpendicular to filaments \citep{kirk2013, fernandez2014} 
have been interpreted as arising from infall onto the filaments (i.e. converging accretion flows). Dust polarization maps show that the plane-of-sky component of the surrounding magnetic field also lies perpendicular to filaments \citep{goodman1992, chapman2011, 2016A&A...586A.136P}. 
That both of these quantities also are aligned with low density `striations' suggests that gas is converging onto filaments, along magnetic fields \citep{palmeirim2013}. Interestingly, fluid motions seem to change direction \textit{inside} of filaments, so that the gas flows \textit{along} filaments, internally \citep{kirk2013, fernandez2014}. Given that the uncertainty of dust polarization measurements increases with density \citep{goodman1992}, the field might actually change direction to become parallel inside of filaments as well. Indeed, if the internal field \textit{did not} become more or less parallel to the filament, external field lines would be bent away from normal due to drag between internal fluid motions and the field (as shown in panel 4A, Fig. \ref{schematic_collapse}). 

The gas motions and accompanying magnetic field geometry associated with filaments have primarily been attributed to magneto-gravitational instability. 
In this conceptualization, an over-dense region within a molecular cloud first becomes Jeans unstable (panel 1A, Fig. \ref{schematic_collapse}). Collapse is triggered and proceeds \textit{along} magnetic field lines, as the gas is not yet magnetically supercritical (panel 2A). Once enough mass accumulates for the region to become magnetically supercritical, gas can begin to fall in perpendicular to the field, as well. For magnetically supercritical filaments, this means collapse would then proceed \textit{along} the filament (panel 3A; \cite{pon2011,pon2012,toala2012}). At this point, the field would become distorted as it is dragged down into the collapsing gas (panel 4A). That the magnetic field strength increases with density in collapsed gas (i.e. $B\propto n,~ n>1000~cm^{-3}$; \cite{troland1986, crutcher1999, crutcher2010, tritsis2015}) has commonly been cited as evidence for this scenario.

An alternative explanation that does not require beginning with a Jeans unstable mass is the reorientation of oblique magnetized shocks, which is the focus of the present paper. This model begins with the collision of marginally supersonic gas along magnetic field lines (i.e. the path of least resistance). In the most general case, where the flows meet at a non-normal interface, it has recently been found that the central shock layer readjusts so that it becomes normal to the upstream field (and oncoming flows). This effect, which has been reported by \cite{kortgen2015} and \cite{fogerty2016}, is also consistent with observed filament gas motions and field morphology (i.e. perpendicular external fluid motions and field lines, as well as parallel internal fluid motions). Moreover, distortion of field lines by passage through colliding flows shocks (cf. \cite{chen2014}) also produces a power law relationship of the magnetic field strength with density \citep{heitsch2007, hennebelle2008, banerjee2009}, consistent with the observations described above. 

To orient the reader, a schematic of the reorientation process is illustrated in the bottom panel of Figure \ref{schematic_collapse}. For the moderately magnetized cases of this paper (those runs that have a $\beta=1$, where $\beta$ is the ratio of the thermal to magnetic pressure), supersonic inflows that meet at an oblique angle ($\theta$) generate both an MHD fast shock ($FS$) and slow shock ($SS$) on either side of a contact discontinuity ($CD$; Fig. \ref{schematic_collapse}B, top/left). Between the $FS$ and $SS$ (region $f-s$), gas that moves parallel to the shock front (note the flow direction is given by the black arrows) drags magnetic field lines \textit{away} from the shock normal, whereas in region $s-s$, field lines bend \textit{toward} the normal as they connect across the $CD$. Over time, the entire post-shock region reorients (right panel), where the grey-shaded region represents the growing filament. As we will show, this process depends heavily on the lateral ejection of material away from the growing filament into the ambient medium (represented by the outward directed arrows, right hand column), which occurs due to pressure gradients between the post-shock gas and the ambient medium. The result is similar when the magnetic field is weakened ($\beta=10$, Fig. \ref{schematic_collapse}B, bottom), notwithstanding the slight changes to the shock structure. 
That is, reorientation produces a post-shock flow and field that has parallel components to the filament, internally (in regions $f-s$ and $f-c$, in the $\beta=1$ and $\beta=10$ cases, respectively), and 
perpendicular components, externally. This MHD shock process is the topic of the present paper. 

We present a suite of 2D magnetized, colliding flows simulations that test the effects of varying the magnetic field strength and the inclination angle of the collision interface on reorientation. We find that reorientation is possible in all but the strongest magnetic field cases, resulting in post-shock filaments\footnote{In 2D, the result is actually the generation of perpendicular \textit{sheets}. Only in 3D could true filaments form. The basic MHD shock physics described here in 2D, however, can be extrapolated to 3D. \cite{fogerty2016} showed that fully 3D simulations also exhibited reorientation.} that approach a normal orientation with respect to the upstream velocity and magnetic field. In addition, we show that fluid motions are highly parallel to the filament within the shock bounded gas, and that oblique, magnetized colliding flows naturally assume an s-shaped structure, reminiscent of the integral-shaped filament \citep{bally1987,johnstone1999}. Our paper is organized as follows. We begin with a description of our numerical methods and model (Section \ref{methods}). We then discuss 1D shock solutions relevant to our work (Section \ref{1dshocksolns}) and our key results (Section \ref{Simulations}). We finish with a discussion of our findings in Section \ref{discussion} and present a resolution study of our results in the appendix.

\section{Methods}\label{methods}

Our simulations were performed using AstroBEAR\footnote{https://astrobear.pas.rochester.edu} \citep{cunningham2009, carroll2013}. AstroBEAR is a massively parallelized, adaptive mesh refinement code designed for astrophysical contexts. The numerical code solves the conservative equations of hydrodynamics and magnetohydrodynamics, and includes a wide-range of multiphysics solvers. A sampling of these solvers include self-gravity, sink particles, various heating and cooling processes, magnetic resistivity, and radiative transfer. The AstroBEAR code is well tested (see, for example, \cite{poludnenko2002,cunningham2009,kaminski2014}), under active development, and fully documented by the University of Rochester's computational astrophysics group.

\begin{figure}
    \centering
    \includegraphics[width=\columnwidth]{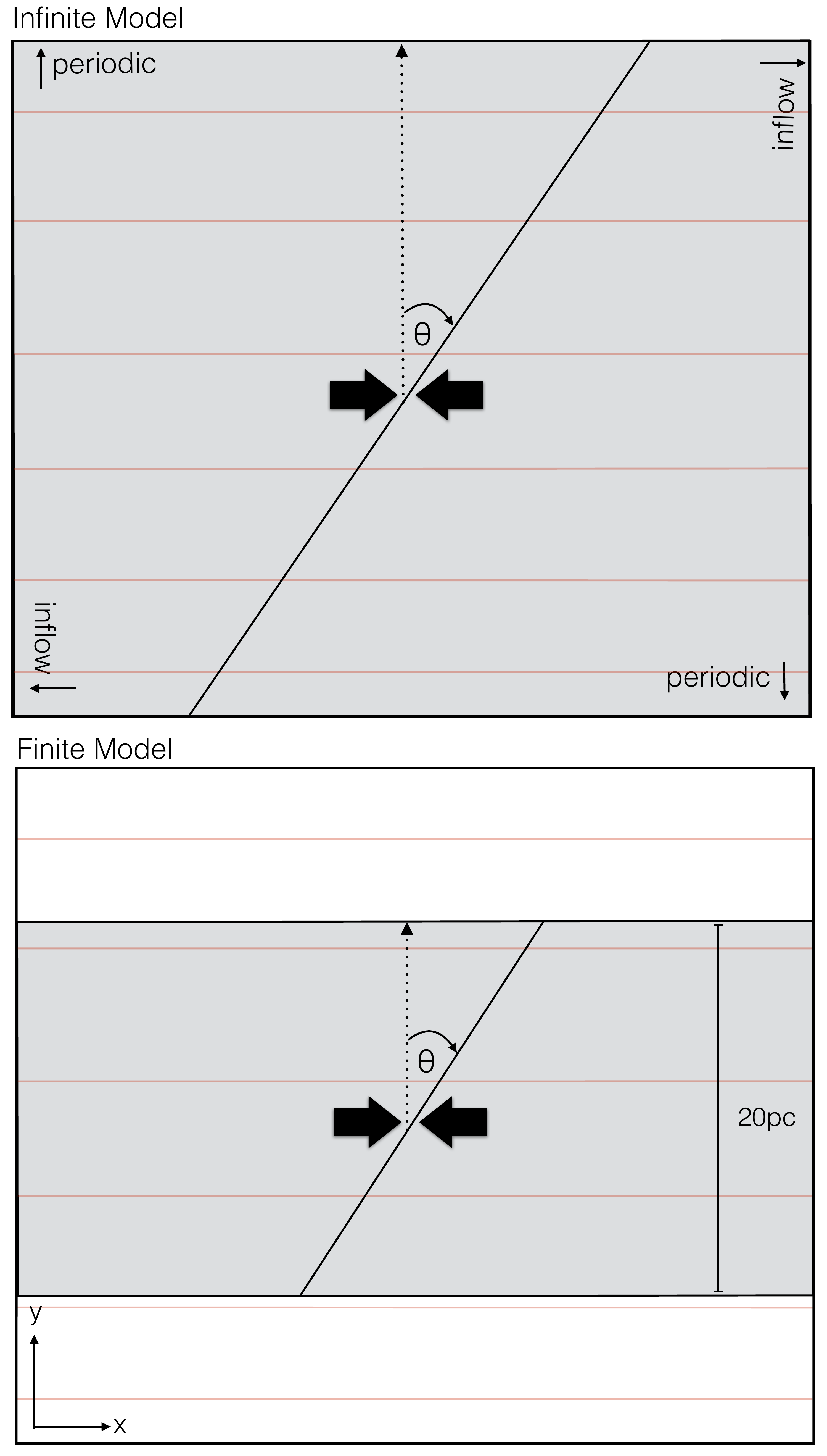}
    \caption{Model diagrams. \textit{Top panel.} Setup for the infinite cases. Incoming flows meet at an effectively infinite oblique shock, defined by the angle $\theta$. Boundary conditions of the different box sides are labeled. Magnetic field lines are represented by the red lines, and the flow is uniform everywhere along the collision interface, in the direction given by the thick black arrows. \textit{Bottom panel}. Setup for the finite cases. The colliding flows are now embedded in a stationary ambient medium of the same density and pressure. The flows are $20~pc$ in diameter and collide at an oblique collision interface, given by $\theta$. The uniform magnetic field is again given by the red lines, and runs  everywhere parallel to the flows.}
    \label{model}
\end{figure}

\begin{table*}
\caption{Suite of simulations.}
\label{tab1}
\begin{tabular}{@{}clclclclclclclclcl@{}}
\hline
Infinite/Finite && Hydro/MHD && $\beta$ && $\theta$ ($\degr$) && Domain Length (pc) && Resolution && AMR levels  \\
\hline
 Infinite$^\dagger$ && Hydro && ... && 30 && 1 && $256^2$ && 0   \\
 Infinite$^\dagger$ && MHD && 1  && 30 && 1 && $256^2$ && 0 \\
 Finite && MHD && .1  && 30 && 160 && $256^2$ && 2 \\
 Finite && MHD && 1  && 30 && 160 && $256^2$ && 2 \\
 Finite && MHD && 10  && 30 && 160 && $256^2$ && 2 \\
 Finite && MHD && .1 && 60 && 160 && $256^2$ && 2 \\
 Finite && MHD && 1 && 60 && 320 && $512^2$ && 2 \\
 Finite && MHD && 10  && 60 && 320 && $512^2$ && 2 \\
\hline
\\
$^\dagger$Did not include cooling
\end{tabular}
\end{table*}

The present suite of simulations consists of two sets of runs. The first was a pair of shock models that reviewed the wave solutions across \textit{infinite} hydrodynamic (hydro) and magnetohydrodynamic oblique shocks, using an exact Riemann solver and HLLD solver, respectively. The shocks were generated by marginally supersonic flows ($M=1.5$, where $M$ is the mach number) that collided at an oblique interface, following \cite{haig2012} and \cite{fogerty2016}. These flows were injected along the $x$ boundaries of a 2D grid so that they completely filled the domain (Fig. \ref{model}, top panel). Boundary conditions in $y$ were set to periodic. The obliquity of the collision interface was given by the inclination angle, $\theta$, and was fixed at $\theta=30\degr$. While these runs were performed on a 2D mesh, the fluid variables varied only along the $x-$dimension. Thus, these runs provide the wave solutions across 1D oblique shocks. The second set of runs was a parameter study of 2D \textit{finite} magnetized colliding flows. In these runs, the flows were embedded in a stationary ambient medium of the same density and pressure (Fig. \ref{model}, bottom panel). The flows were again marginally supersonic ($M=1.5$), and collided at an oblique interface, given by the inclination angle $\theta$. 
The inclination angle of the finite runs varied between $\theta=30\degr$ and $60\degr$. The complete suite of simulations is given in Table \ref{tab1}. 

Self-gravity was not included in the present suite of simulations. Consequently, the simulations neglected gravitational collapse along filaments. The runs instead tracked the formation and early evolution of filaments, while illustrating the basic mechanism of reorientation. 
Cooling was included in all of the finite colliding flows runs, following a modified \cite{inoue2008} cooling curve. The modification allowed the gas to cool to $T=10~K$, appropriate for ISM conditions (Ryan \& Heitsch, in prep). The infinite cases did not include cooling, but instead used an adiabatic equation of state with $\gamma=5/3$.

The parameters of each of the runs were chosen to match ideal ISM conditions (as discussed in \cite{fogerty2016}). Thus, the runs were initialized at a uniform number density of $n=1 ~cm^{-3}$ and temperature of $T=4931~K$. At these densities and temperatures, the gas was initially in thermal equilibrium (for those runs that included cooling). The speed of each inflow was $v=11~km~s^{-1}$. 
For those runs that included a magnetic field (cf. Table \ref{tab1}), the field was uniform throughout the simulation domain and ran parallel to the flows. The strength of the field was set by $\beta$. In the 1D MHD case, $\beta=1$. For the finite colliding flows runs, $\beta$ ranged between  $\beta=10-.1$. This is equivalent to a field strength of $B=1.6-16~ \mu G$, in line with current measurements of the global mean field of the ISM ($B\approx1-10 ~\mu G$, \cite{beck2001, heiles2005}).

The infinite colliding flows runs had periodic boundary conditions in $y$ and inflow boundary conditions in $x$. The length of the square 2D domain was $1~pc$ on a side. The mesh was a fixed-grid at a resolution of $256^2$ and had a cell size of $\triangle x=.0039~pc$. For the finite colliding flows runs, the size of the square domain was chosen to be large enough so that no gas or magnetic field lines left the box over the course of the simulation (cf. Table \ref{tab1}). However, the effective resolution was held constant at a finest cell size of $\triangle x_{min}=.15625~pc$. Boundary conditions were set to inflow where the flows were injected, and extrapolating everywhere else. The radius of the finite colliding flows was $r=10~pc$. 
The final simulation time for the finite runs was $t_{sim}=12 ~Myr$. The infinite cases were shorter at $t_{sim}=1~Myr$, the  time it took for the shocks to reach the edge of the simulation domain. 

\section{Infinite Adiabatic Oblique Shocks}\label{1dshocksolns}

Before we present the simulations of reorienting MHD colliding flows, we briefly discuss the hydro and MHD shocks that are generated by infinite, adiabatic colliding flows. The results of this section provide a framework for evaluating the shocks that are formed between finite, cooling colliding flows. Given this section is just a review of 1D oblique shock solutions, the reader can feel free to jump ahead to our main results in Section \ref{Simulations}.

\subsection{Hydro Case}\label{hydro1d}

The hydro Riemann problem relevant to the present paper consists of a constant density and pressure fluid that is separated by a discontinuous jump in velocity. In particular, the velocity field converges on the central Riemann interface at an oblique incidence. Note, the inclination angle of the interface for both the infinite hydro and MHD run (discussed below) was chosen to be $\theta=30\degr$, to match the finite $\theta=30\degr$ runs of Section \ref{Simulations}.

\begin{figure}
    \includegraphics[width=\columnwidth]{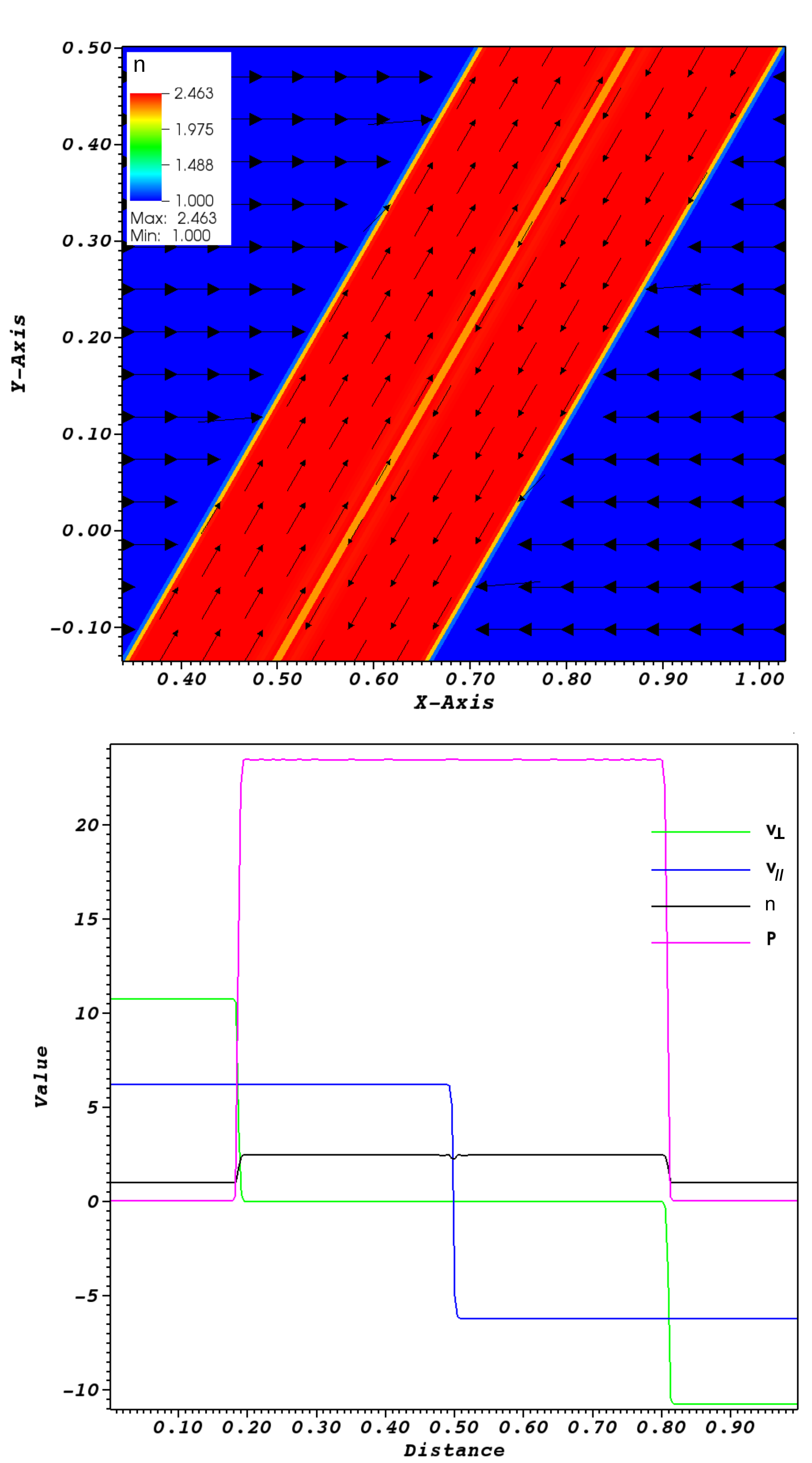}
    \caption{The fluid variables across hydrodynamic, oblique shocks. \textit{Top panel} shows number density of the infinite hydrodynamic Riemann problem described in text. Velocity vectors are overlaid and scaled by magnitude. 
    \textit{Bottom panel} gives the various fluid variables across the waves. Perpendicular and parallel components of velocity ($v_\perp$ and $v_{//}$, respectively) are in units of $km ~s^{-1}$, number density in $cm^{-3}$, and pressure ($P$) in $K~cm^{-3}$. Note, pressure has been scaled to fit on the $y-$axis, where $P=(P-4931)\times10^{-3}$.}
    \label{PS_hydro}
\end{figure}

As can be seen in Figure \ref{PS_hydro}, this Riemann problem generates two oblique shocks that are separated by a contact discontinuity. In addition to the characteristic density increase across each shock front, the top panel of Figure \ref{PS_hydro} shows that velocity vectors bend away from the shock normal across each shock. This occurs because only perpendicular components of upstream velocity vectors ($v_\perp$) change across shocks. 
This leads to $v_\perp \rightarrow 0$ in the downstream gas. Note, if $v_{\perp} \neq 0$ in the post-shock gas, additional waves would be generated behind the shocks, which would violate the three-wave solution family of hydrodynamic Riemann problems. Thus, post-shock gas flows \textit{exactly} parallel to each shock front. In other words, \textit{oblique shocks generate shear}. The various fluid variables across the wave modes are given in the bottom panel of Figure \ref{PS_hydro}.

\subsection{MHD Case}\label{1dmhd}

\begin{figure}
    \includegraphics[width=\columnwidth]{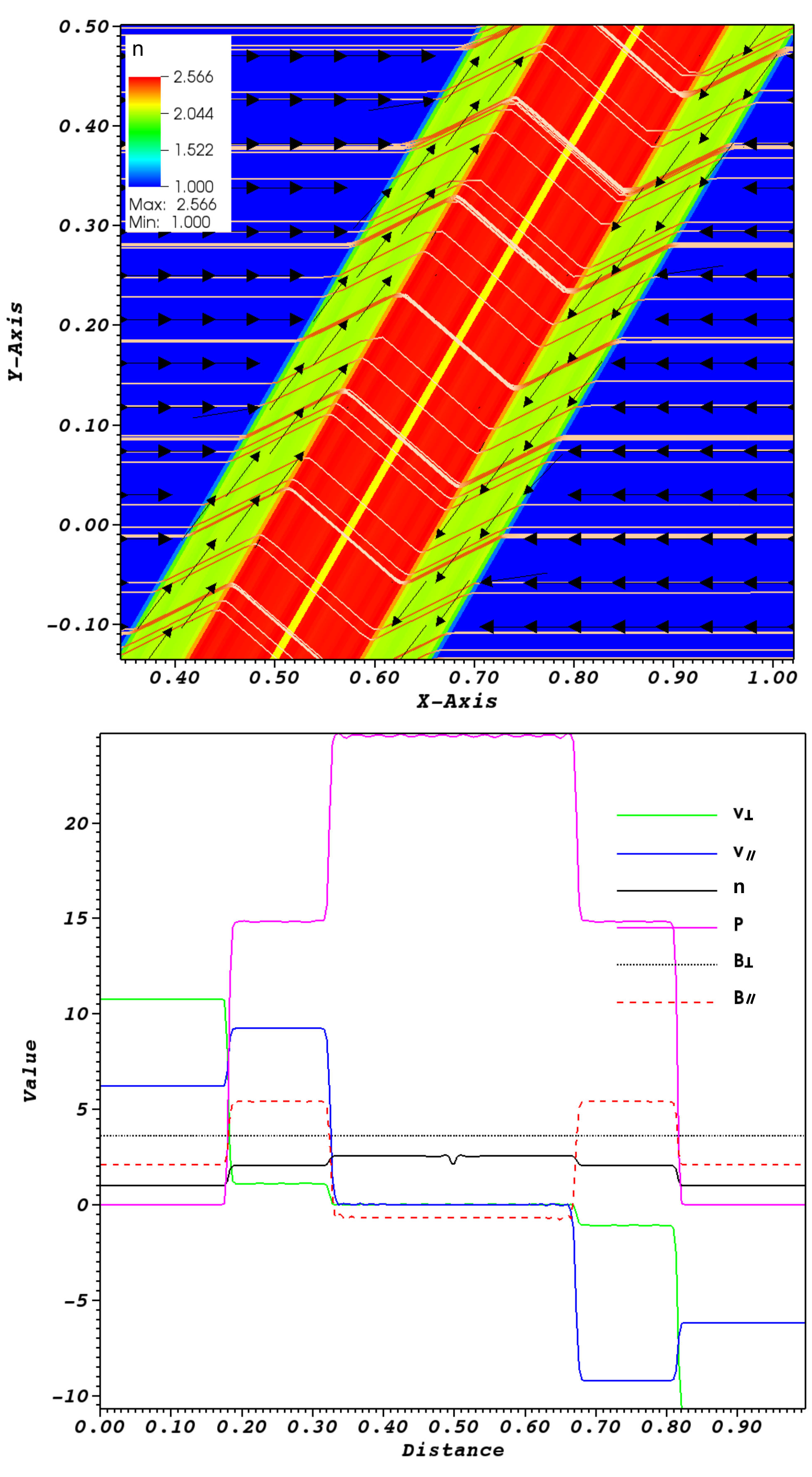}
    \caption{The fluid variables across MHD, oblique shocks.\textit{Top panel} shows number density of the infinite MHD Riemann problem described in text. Field lines increase in strength from white to red, and velocity vectors are scaled by magnitude. \textit{Bottom panel} gives the fluid variables across the waves. Units are the same as in Figure \ref{PS_hydro}, with the addition of the perpendicular and parallel magnetic field components ($B_\perp$ and $B_{//}$, respectively), given in $\mu G$.}
    \label{1dMHDpseudo}
\end{figure}

The addition of a uniform magnetic field that is parallel to the flows modifies the shock structure just described. For this case, {\it two} oblique shocks are generated on either side of the contact discontinuity (Fig. \ref{1dMHDpseudo}). Given Alfv\'{e}n modes cannot be generated in 1D, 
the forward-most shock can be identified with the MHD fast shock ($FS$), which is trailed by the slow shock ($SS$). As before, incoming velocity vectors are deflected away from the shock normal across the outer, $FS$ (Fig. \ref{1dMHDpseudo}). However, now these redirected velocity vectors encounter the magnetic field. The collision between the flow and the field causes the field to also bend away from the shock normal across the FS, as the field and fluid are perfectly coupled in ideal MHD. 
The result is an amplification of the field across the $FS$, visible by both the color of the field lines, which increase in strength from white to red (Fig. \ref{1dMHDpseudo}, top panel), as well as the amplification of the parallel field component ($B_{//}$) across the $FS$ (bottom panel). Given the field is tied across the contact discontinuity ($CD$), it eventually must turn back toward the shock normal. This gives rise to the inner, $SS$, where the field again switches direction. However, in contrast to the $FS$, gas that passes the $SS$ is stagnated. This is visible by zero velocity gas motions in the region $s-s$ (defined in Fig. \ref{schematic_collapse}, bottom panel). Lastly, we note that reorientation fundamentally cannot occur in 1D, given the symmetry in the initial conditions. 

\section{Finite Oblique Shocks, with cooling}\label{Simulations}

We now turn to our finite magnetized colliding flows runs to address the issue of reorientation of oblique shocks. Note, this section differs from the last in that now the flows are embedded in a stationary ambient medium. We begin by discussing the effects of varying $\beta$ and $\theta$ on the morphology of reoriented flows. We then move on to discussing the temporal evolution of each of the cases, again focusing on morphological changes over the course of the simulations.

\subsection{The Effects of Varying $\beta$ and $\theta$ on Reorientation}

\begin{figure*}
    \includegraphics[width=.86\textwidth]{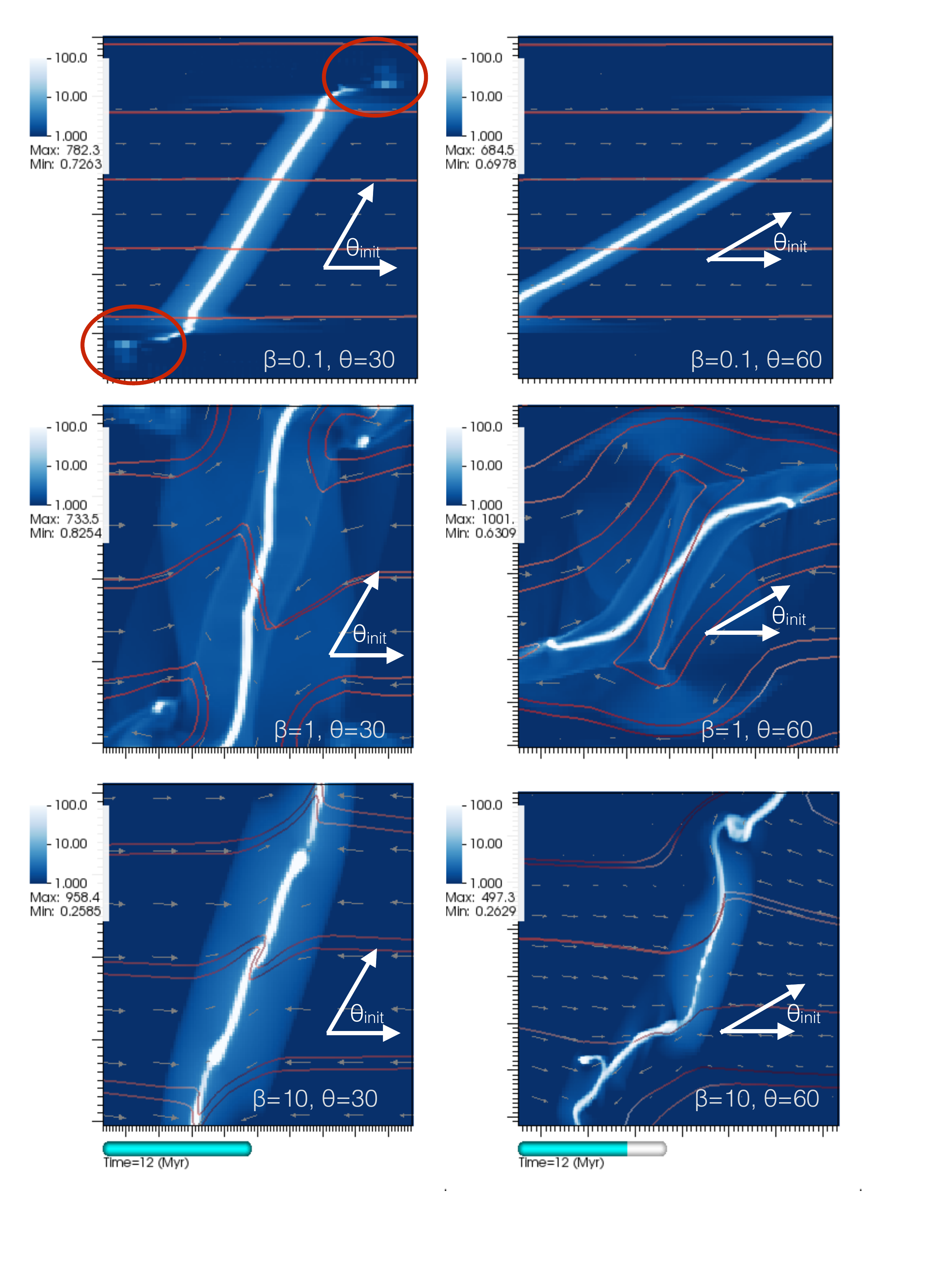}
    \caption{Density plots of the six finite colliding flows runs, with overlaid velocity vectors and magnetic field lines. Each plot shows the flows at the final simulation time of $12~Myr$. The legend gives number density in units of $cm^{-3}$. Vectors are scaled by magnitude, and field lines increase in strength from white to dark red. The initial inclination angle of the collision interface is given by $\theta_{init}$. The trailing arms of the filament (discussed in text) are illustrated by the red circles in the upper-left panel. The location of these trailing arms extend beyond the plot boundaries for the weaker field cases.}
    \label{fieldlines_comparison}
\end{figure*}

In the finite colliding flows scenario, post-shock motions no longer strictly adhere to the shear flow predicted by the infinite shock solutions of Section \ref{1dshocksolns}. Pressure gradients now exist between post-shock gas and the external ambient medium, which forces material laterally outwards, away from the collision region. Given reorientation does not occur in the infinite case, this lateral motion is a key component of reorientation. This is illustrated by the fact that reorientation does not occur when material is \textit{prevented} from leaving the collision region, as in the strong field ($\beta=.1$) cases. Plots of number density show that when $\beta=.1$, for inclination angles of either $\theta=~30\degr$ or $60\degr$, magnetic field lines are strong enough to resist ejection of material from the collision region  (Fig. \ref{fieldlines_comparison}, top panel). Any material that does manage to escape the collision region travels parallel to the colliding flows, along the field lines, and is deposited in what we call the `trailing arms' of the filament -- a prominent feature of finite, oblique, magnetized colliding flows (see red circles in Fig. \ref{fieldlines_comparison}, top-left panel). Indeed, the field is so strong in these cases that a post-shock shear flow is also inhibited, as the field effectively resists motion perpendicular to it. Instead, post-shock material is delivered directly onto the growing filament, where it continues to collect, cool, and compress over the course of the simulation. This results in a smooth and flat filament, as turbulent sub-structure (induced from the thermal instability, e.g.) is also effectively inhibited.

The behavior begins to change as the magnetic field is weakened. When $\beta$ is increased to $1$, the $\theta=30\degr$ case exhibits a large-scale reorientation. This is illustrated in Figure \ref{fieldlines_comparison} (middle row, left panel), where we see an initially inclined filament has reoriented to become more or less normal to the oncoming flows. As in the infinite version of this case, an outer shock layer forms that diverts incoming flows either diagonally `up' or `down', depending on the upstream interface orientation (cf. Section \ref{1dmhd}).
However, across the second, inner shock (region $s-s$), gas is falling onto the long axis of the filament, along magnetic field lines. This is in contrast to the infinite case, where fluid motions were absent in this region, and thus, these motions are arising from cooling. The result is the formation of a post-shock flow that has parallel components to the filament in both the $f-s$ and $s-c$ regions. 

Near the top and bottom of the filament (i.e. near the flow/ambient boundary), and across the $CD$, the velocity vectors become aligned so that they are pointing along $y$, out into the ambient medium. That is, the flow switches from being a shear flow across the $FS$, to being directed outwards into the ambient medium in the entire $f-f$ region (near the flow/ambient boundary). This is due to pressure gradients between the collision region and the ambient medium. The ejection of material from the over-pressurized collision region into the ambient medium drives arcs in the magnetic field, whose tension impedes further lateral flow and traps the ejected gas (see \cite{fogerty2016} for an analytical description of this process). This trapped gas travels along the magnetic field line arcs, as shown by the corresponding velocity vectors in Figure \ref{fieldlines_comparison}, where it collects in the trailing arms. This three-step process, namely, 1) post-shock ejection from the collision region into the ambient medium, 2) the bending of the magnetic field lines into arcs, and 3) the redirection of incoming flows along the arcs to collect in the trailing arms of the filament, produces an `s-shaped' filamentary structure. 

As the inclination angle is steepened to $60\degr$ for the $\beta=1$ case, the same trends develop. Namely, material is deflected away from the shock normal across the $FS$, ejected from the collision region into the ambient medium, and an s-shaped filament forms with a similar degree of reorientation by $t=12~Myr$. 
While the higher obliquity shocks of this run produce weaker post-shock compression and thus a wider $f-s$ region, changing the inclination angle does not appear to drastically alter the degree of reorientation for the moderate field, $\beta=1$ cases.

As the field is weakened further in the $\beta=10$ runs, the $30\degr$ case reorients to a similar degree, but reorientation is enhanced in the $60\degr$ case (Fig. \ref{fieldlines_comparison}, bottom row). We also see that slightly different filamentary structures form, as there are no longer slow shocks present in the filaments. This means that by $\beta=10$, the shock solutions have effectively approached the 1D hydro solutions (away from the flow/ambient boundary), as they must, in the limit of weak magnetic field. 
Furthermore, there are no longer $s-s$ regions (marked by a stagnate velocity field), where material can expand away from the collision region due to pressure gradients alone. Instead, incoming gas is quickly shunted away from the collision region as it passes through the single oblique shock on either side of the $CD$. As this material leaves the collision region, it drags magnetic field lines away with it. As the plots show, the weaker field is dragged to further distances, and thus the trailing arms extend farther away into the ambient medium from the main filament body (beyond the plot boundaries). 

\subsection{Temporal Evolution of the Flows}

\begin{figure*}
    \includegraphics[width=\textwidth]{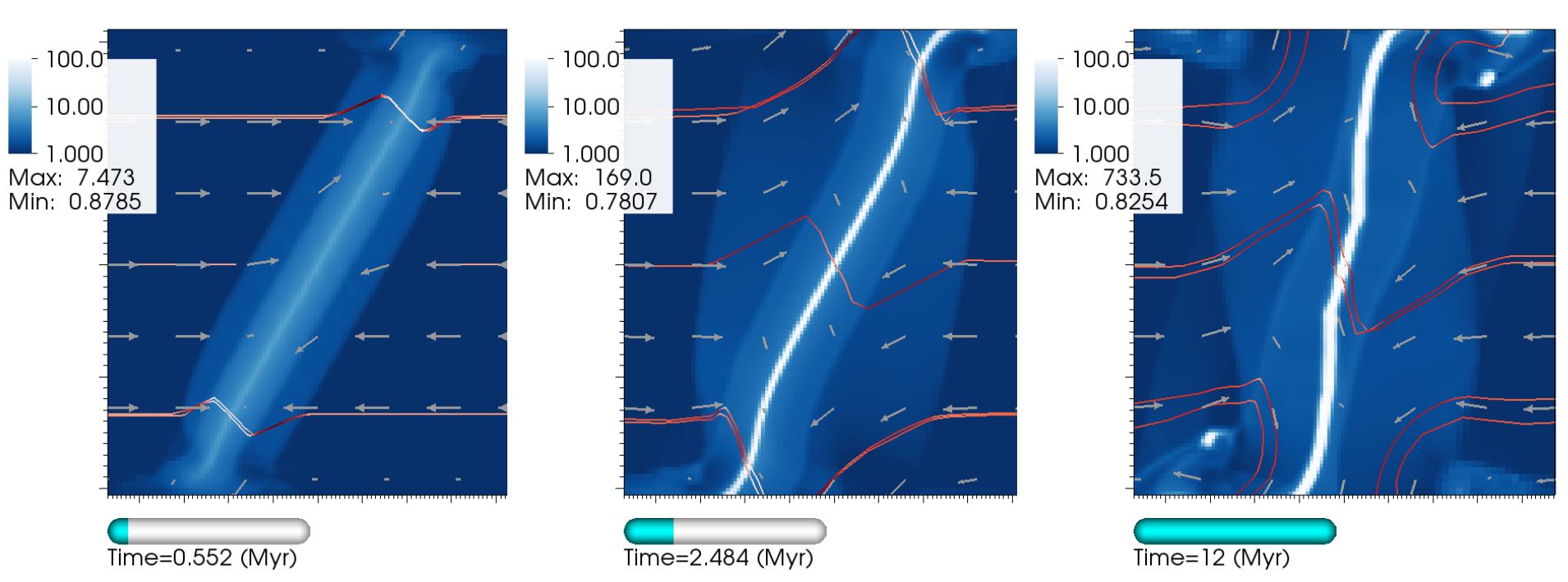}
    \caption{Temporal evolution of the $\beta=1,~\theta=30\degr$ case. All units are the same as in Figure \ref{fieldlines_comparison}.}
    \label{s30b1progression}
\end{figure*}

\begin{figure*}
    \includegraphics[width=\textwidth]{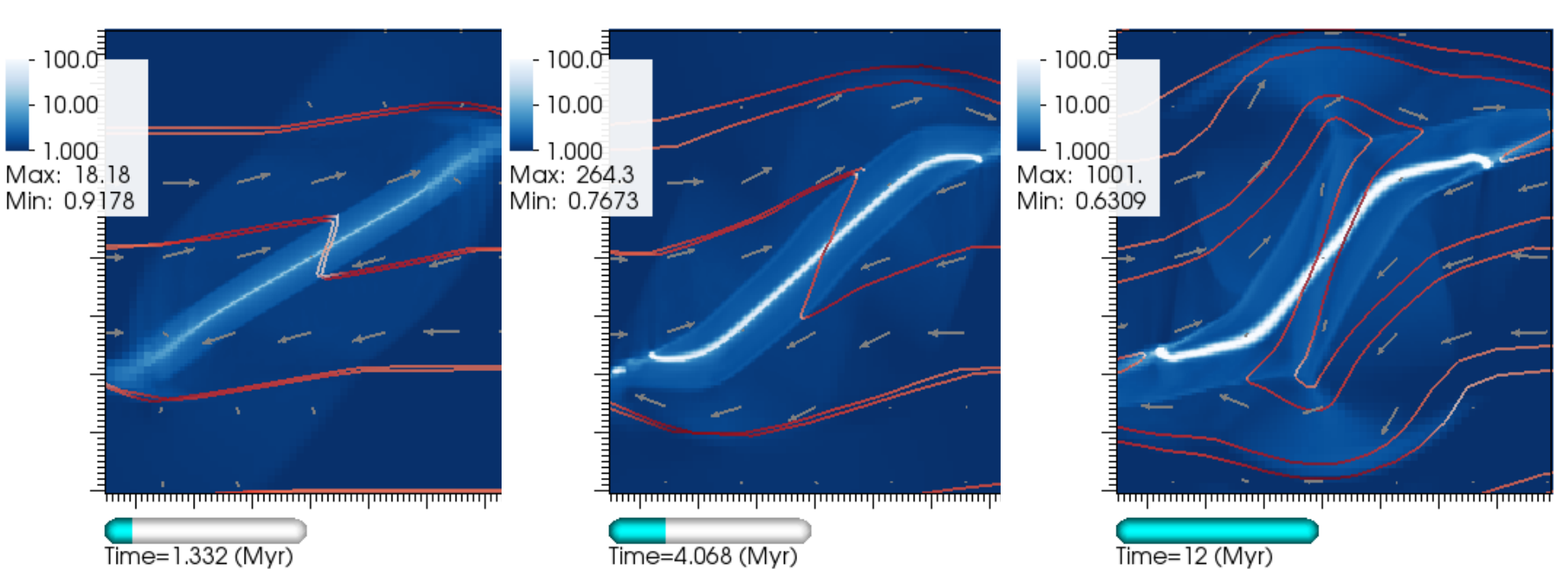}
    \caption{Temporal evolution of the $\beta=1,~\theta=60\degr$ case. All units are the same as in Figure \ref{fieldlines_comparison}.}
    \label{s60b1progression}
\end{figure*}

We now turn to the temporal evolution of the flows. We will again be focusing on density plots with overlaid velocity vectors and magnetic field lines, beginning with the stronger field ($\beta=1$) cases. Figure \ref{s30b1progression} shows the evolution of the $\beta=1, ~\theta=30\degr$ case. As can be seen in the $t=.55~Myr$ panel, the bulk properties of the flow are similar to the infinite MHD case (Section \ref{1dmhd}), at early times. That is, a shear flow is established across the outer, $FS$, and negligible fluid motions occur inside of region $s-s$. Over time, post-shock cooling triggers inflow along the field lines onto the $CD$ of the forming filament ($t=~2.5~Myr$). Additionally, this time panel shows that the fast shocks stall (near the top flow/ambient boundary for the right $FS$, and near the bottom for the left $FS$). This is due to a decrease in thermal pressure support behind the shocks as material escapes into the lower pressure ambient medium. The result of these stalled shock fronts is the reorientation of the outer shock layers, as those regions that are not stalled  continue to propagate away from the collision region. This leads to the outer shock layers becoming normal to the oncoming flows. 

As can be seen in the figure (and even better in the animations online\footnote{See the following youtube channel: https://www.youtube.com/channel/UCE mUg0BdCyPC3QnKdNDJq1w}), reorientation of the inner shock layer (i.e. region $s-s$) also begins with the lateral ejection of material from the collision region. That is, reorientation occurs from the \textit{outside-in} -- beginning near the flow/ambient boundary and moving toward the colliding flows axis by $t=12 ~Myr$. This occurs as material is delivered along the `z-shaped' magnetic field lines near the colliding-flows axis nearly vertically onto the long axis of the filament. When the inclination angle is steepened to $\theta=60\degr$ for the same value of $\beta=1$, the evolution is qualitatively similar (Fig. \ref{s60b1progression}).

\begin{figure*}
    \includegraphics[width=\textwidth]{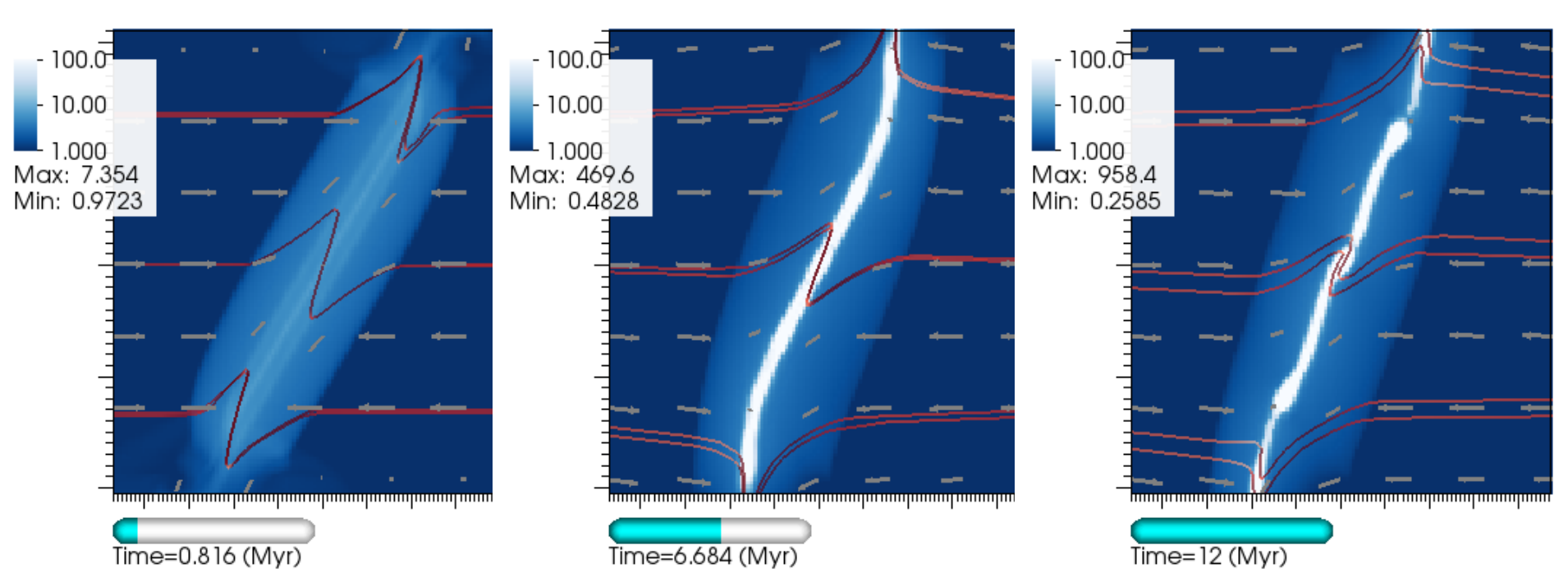}
    \caption{Temporal evolution of the $\beta=10,~\theta=30\degr$ case. All units are the same as in Figure \ref{fieldlines_comparison}.}
    \label{s30b10progression}
\end{figure*}

\begin{figure*}
    \includegraphics[width=\textwidth]{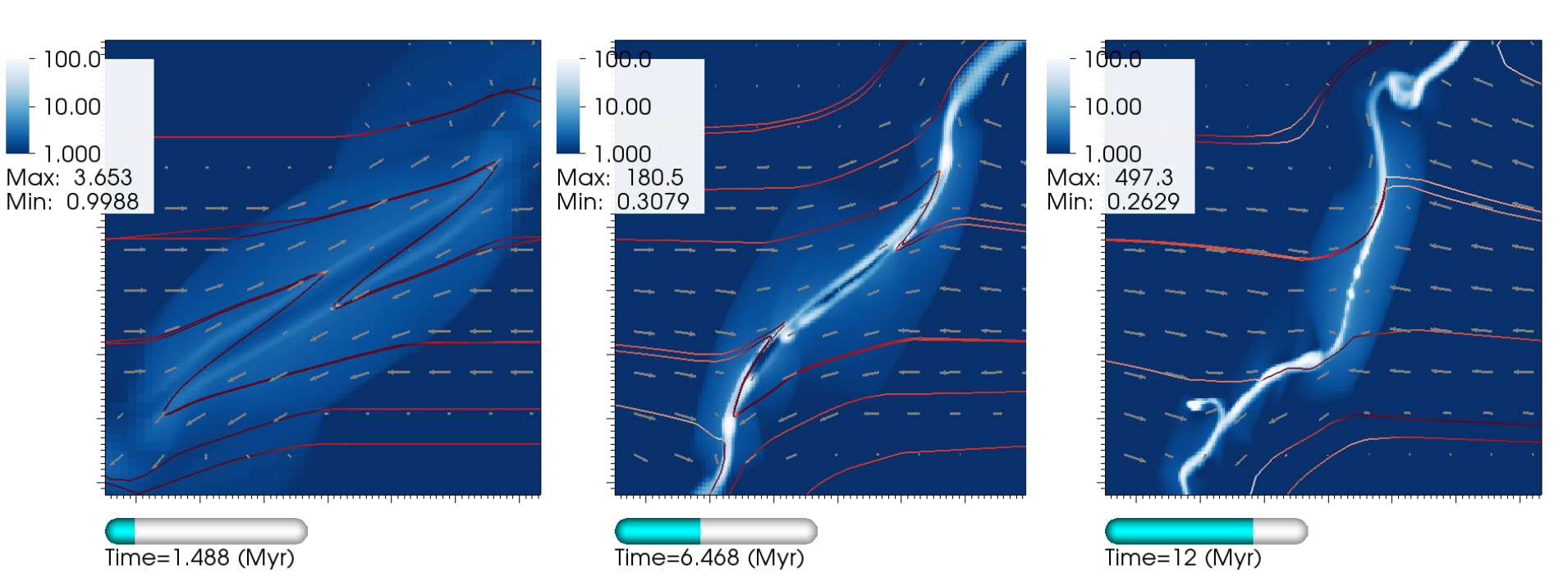}
    \caption{Temporal evolution of the $\beta=10, ~\theta=60\degr$ case. All units are the same as in Figure \ref{fieldlines_comparison}.}
    \label{s60b10progression}
\end{figure*}

As the field is weakened to $\beta=10$, the largest difference that occurs in the evolution is the appearance of a vacated inner region along the filament's long axis (FLA) as material preferentially collects around the top and bottom of the filament (i.e. near the flow/ambient boundary), as well as just exterior to the FLA (see Figures \ref{s30b10progression} and \ref{s60b10progression}, middle panels). This arises from the extreme amplification of the field along the $CD$ as field lines are stretched by the shear flow (recall, the color of the field lines increases from white to dark red with increasing field strength). By the last time panel, in both the $\theta=30\degr$ and $60\degr$ cases, gas has successfully collapsed onto the FLA, and the internal high density regions of the filament have begun to reorient, again from the outside-in. Within the region $f-f$, the flow and the field are highly parallel to the filament, and the entire structure (filament+trailing arms) has assumed an s-shaped geometry.

While this overall evolution is the same between the two $\beta=10$ cases, the external flow field in the $\theta=60 \degr$ case at $t=12~Myr$ is unlike any of the other runs. Velocity vectors show that material entering from the left is directed downwards as it approaches the filament, and that material entering from the right is directed upwards. Note, this is the \textit{opposite} flow pattern that the oblique shocks establish alone (see earlier time panels), indicating that the flow is traveling along bowed magnetic field lines. Such an asymmetric flow is necessary to generate a torque on the filament. This could explain why the $\beta=10,~\theta=60\degr$ case reorients more than the $\beta=1,~\theta=60\degr$ case (see also Fig. \ref{reorientation}).

\begin{figure*}
\centering
    \includegraphics[width=.8\textwidth]{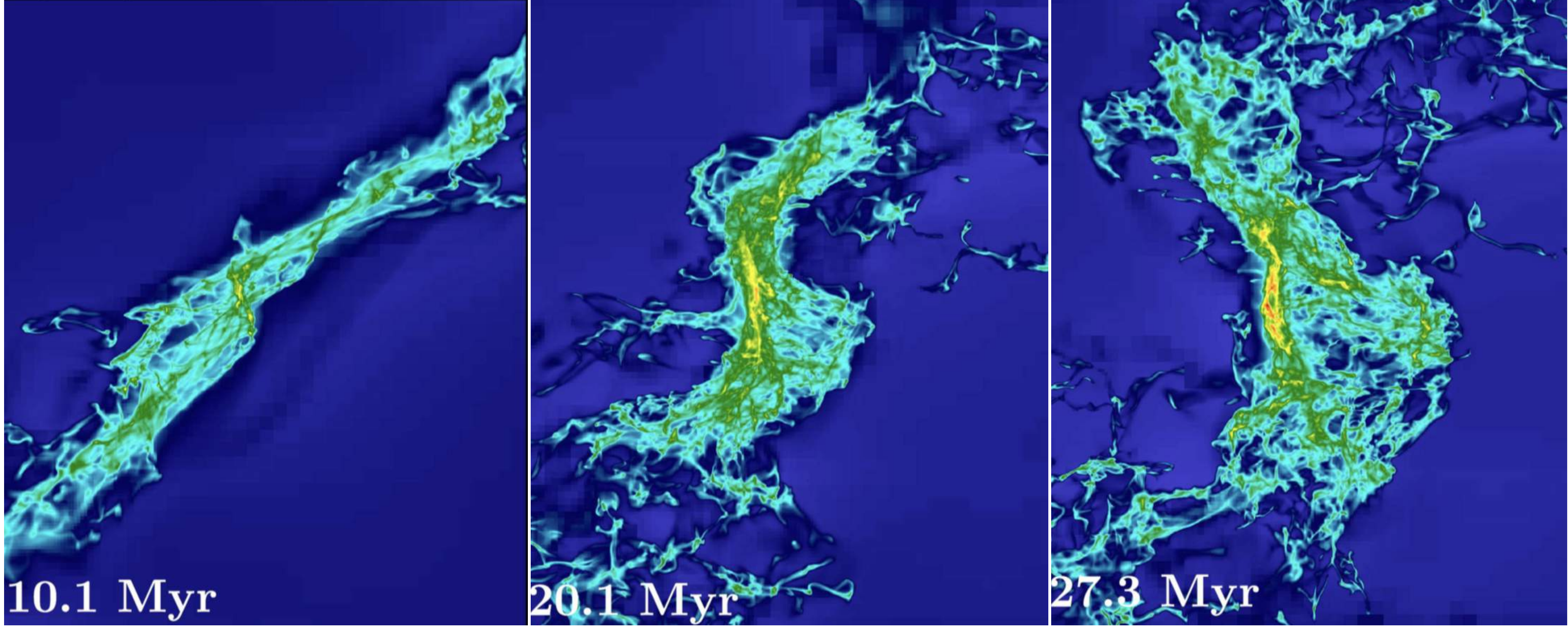}
    \caption{Evolution of the 3D $\beta=10,~\theta=60\degr$ case, with self-gravity. Image taken from \protect\cite{fogerty2016}. The colors represent column density, and increase from blue to red. See \protect\cite{fogerty2016} for details.}
    \label{3Drun}
\end{figure*}

\section{Discussion}\label{discussion}

We have presented an MHD shock mechanism capable of reorienting filaments formed via colliding flows. This process was identified previously in fully 3D simulations of magnetized colliding flows that included turbulence, self-gravity, cooling, and oblique shocks at the collision interface \citep{fogerty2016}.  In that work, large-scale reorientation of molecular gas had occurred for a highly oblique run (Fig. \ref{3Drun}). Reorientation of the collision interface of 3D, magnetized, oblique colliding flows was also recently reported by \cite{kortgen2015}. By running the simulations in 2D, without gravity, we have shown that the mechanism of reorientation is robust. 
Moreover, that hydro simulations of oblique colliding flows do not exhibit reorientation (both in 2D, as well as 3D, \cite{haig2012}, Haig \& Heitsch, in prep), indicates that magnetic fields play a crucial role in the process. 

As we have shown, reorientation begins with the lateral ejection of material away from the post-shock collision region between colliding flows. 
Without magnetic fields, ejecta would continue to push out into the ambient medium until its ram pressure came into balance with the thermal pressure of the environment. With magnetic fields, this material is funneled back down along bowed magnetic field lines, toward the colliding flows surface. This produces what we have called, `trailing arms' of the filament. Additionally, escaped gas from the collision region into the ambient medium translates into a loss of post-shock pressure support, resulting in stalled outer shocks. This led to the reorientation of the outer shocks, which \textit{is} also seen in hydro. The difference lies in the reorientation of the densest regions of the filament (i.e. along the $CD$ of the collision region), which does \textit{not} occur without the magnetic field. Instead, the shear flow setup across hydro oblique shocks generates instabilities (i.e. KH modes) along the interface. These instabilities in turn produce  a `stair-casing' structure, but not a coherent filament that reorients \citep{haig2012}. \cite{hennebelle2013} showed that in order to build a coherent filament in a shear flow, a magnetic field must be present.

\begin{figure*}
\centering
    \includegraphics[width=.8\textwidth]{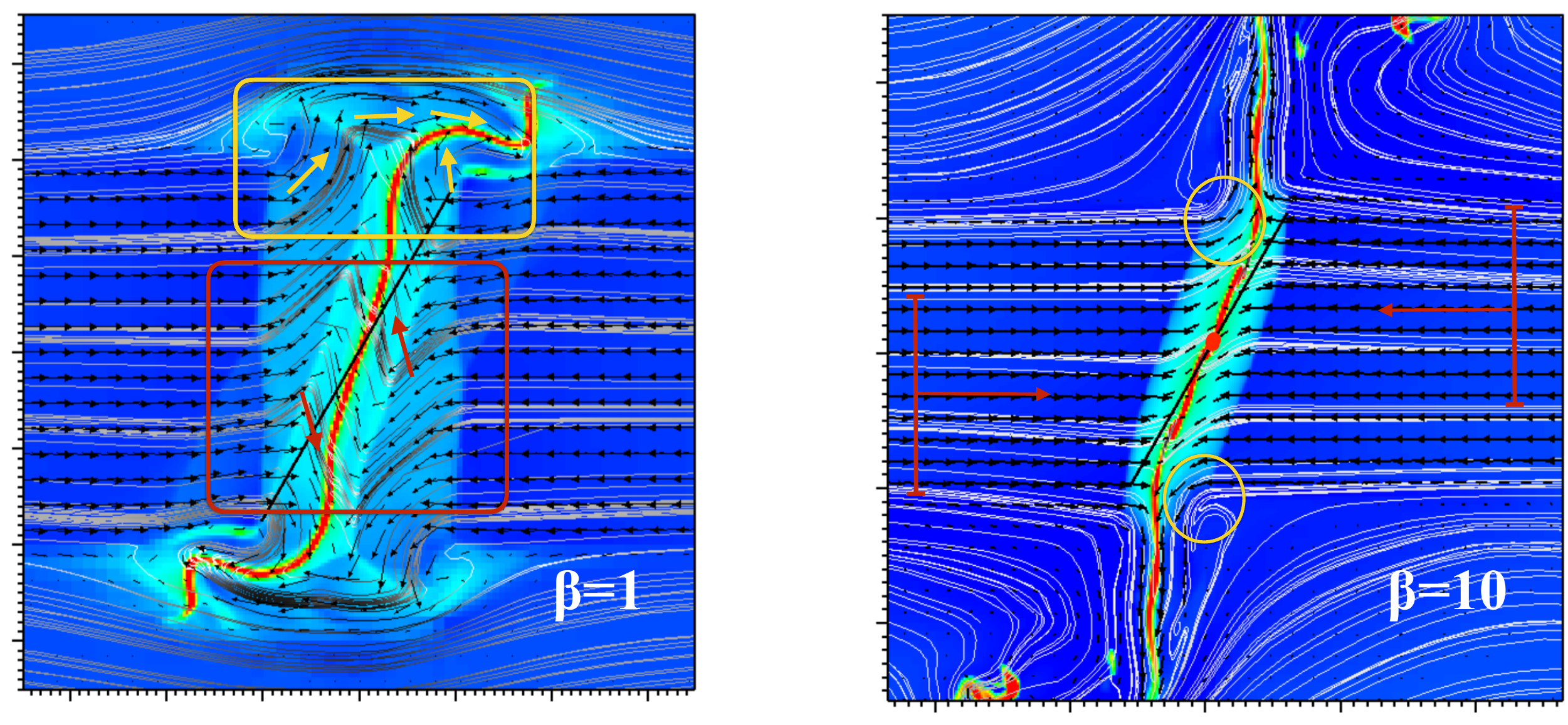}
    \caption {Potential reorientation mechanisms for finite, oblique, MHD colliding flows. \textit{Left panel} shows the dominant mechanism of reorientation for the $\beta=1$ cases. Bowed field lines near the flow/ambient boundaries (e.g. yellow box) deliver material away from the collision region onto the trailing arms, whereas `z-shaped' field lines nearer the colliding flows axis (red box) deliver material directly onto the filament. This differential delivery results in an outside-in reorientation of the filament. \textit{Right panel} shows that in the weaker field cases, material escapes from the collision region asymmetrically across the $CD$. In regions marked by yellow circles, the post-shock shear flow is \textit{enhanced} as material flows out of the collision region along the flow/ambient pressure gradient. In contrast, for oppositely directed material across the shock layer, this pressure gradient now \textit{counteracts} the shear flow. The result is a mismatch of incoming $x-$momentum onto the filament (red, barred arrows) and a torque in the direction of reorientation.}
    \label{reorientation}
\end{figure*}

The role of the magnetic field in reorienting the internal layers of the filament is the delivery of material along `z-shaped' magnetic field lines within the collision zone, near the colliding flows axis (red box, Fig. \ref{reorientation}, left panel). Note, these field lines have not been `blown-out' into arcs from the lateral post-shock ejection (as seen in the yellow box of Fig. \ref{reorientation}, left panel), and thus deliver material nearly vertically onto the growing filament. Thus, reorientation begins with the lateral escape of gas from the collision region, and moves toward the center of the collision region over time. As the magnetic field is weakened, torque on the collision region may play an additional role in the reorientation process. As we have seen, the $\beta=10,~\theta=60\degr$ case reoriented to a greater extent than the stronger field $\beta=1, ~\theta=60\degr$ case, and this coincided with a large-scale asymmetric flow onto the filament. 
The mismatch in incoming $x$-momenta onto the filament can be seen in the right-hand panel of Figure \ref{reorientation}, which illustrates the effect for the smaller inclination angle case, $\beta=10,~\theta=30\degr$. As the figure shows, there is an asymmetric loss of material from the collision region across the $CD$ in the weak field cases (yellow circles), which could lead to a net torque being generated by the bounded flow that remains (red arrows).

Lastly, we have demonstrated that reorientation produces structures that are reminiscent of those found in nearby star-forming regions. Namely, we have shown that under realistic ISM magnetic field strengths, reorientation generates filaments perpendicular to their background magnetic fields, as well as exhibit velocity motions that switch from being perpendicular externally to parallel internally. These features are consistent with observations of star-forming filaments \citep{kirk2013,palmeirim2013,fernandez2014,2016A&A...586A.136P}. Additionally, we have shown that filaments formed via finite, magnetized oblique shocks naturally assume an s-shaped geometry.  

\section{Appendix: Resolution Study}\label{appendix}

\begin{figure*}
\centering
    \includegraphics[width=\textwidth]{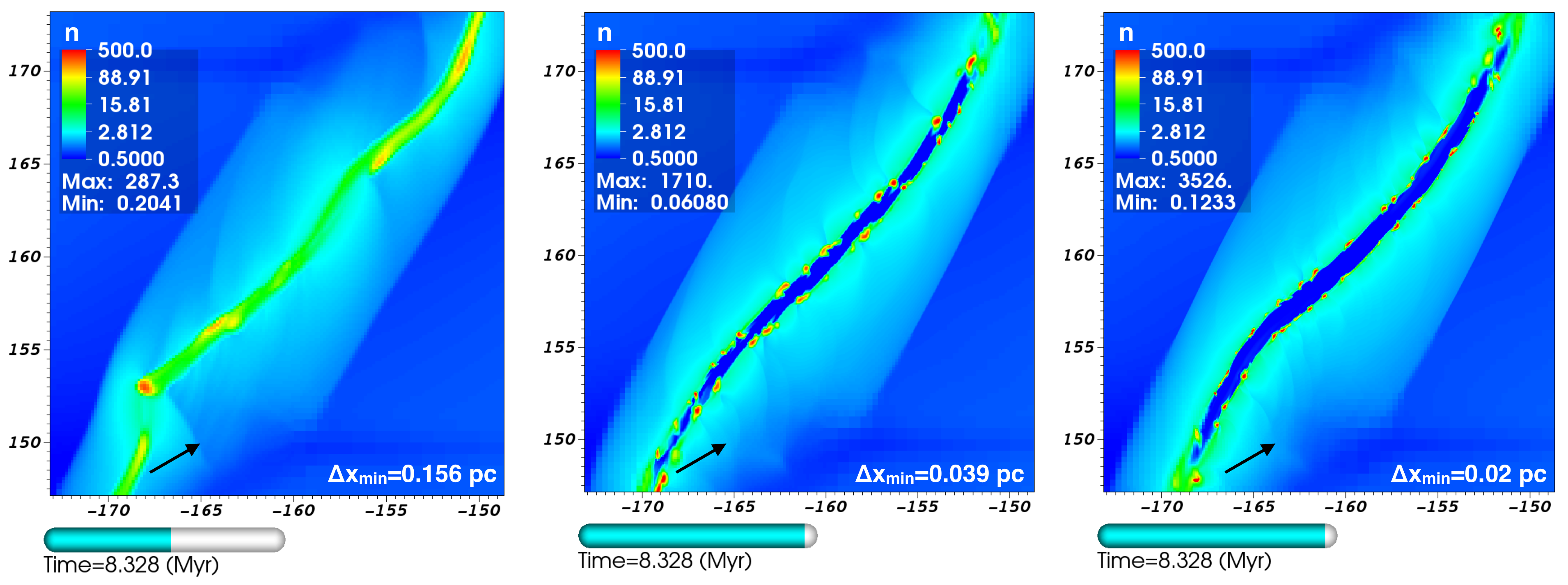}
    \caption {Resolution study of the $\beta=10,~\theta=60\degr$ case. The three panels show a zoom-in of the collision region of this finite case at $8.3 ~Myr$ for three test resolutions: $\Delta X_{min}=0.156, ~0.039,$ and $0.02~pc$. The leftmost panel is the resolution of the current paper. The color legend corresponds to number density ($cm^{-3}$), and the black arrow shows the initial orientation of the shock layer.}
    \label{convergencestudy}
\end{figure*}

We conducted a short resolution study of the $\beta=10$, $\theta=60\degr$ case, which shows that increasing the resolution to a minimum cell size of 0.02 pc (3 additional AMR levels), \textit{does not change the reorientation behavior of the MHD oblique shock layer}. As Figure \ref{convergencestudy} shows, however, a couple of differences are present at higher resolution. First, the internal structure of the filament appears to not be fully resolved in the present paper: a low-density, internal layer of the filament is present at higher resolution (Fig. \ref{convergencestudy}, middle and right panels). Note, this structure is present early-on, at lower resolution (c.f. leftmost panels of Figures \ref{s30b10progression} $\&$ \ref{s60b10progression}). This suggests that numerical diffusion diminishes magnetic pressure in this region in the lower resolution runs, thereby allowing material to cool and condense onto the axis. Indeed, as the resolution increases, the total magnetic energy in the simulations continues to increase as well (for $t>2.4 ~Myr$, Fig. 13), as post-shock compressions result in magnetic field amplification.

\begin{figure}
\centering
    \includegraphics[width=\columnwidth]{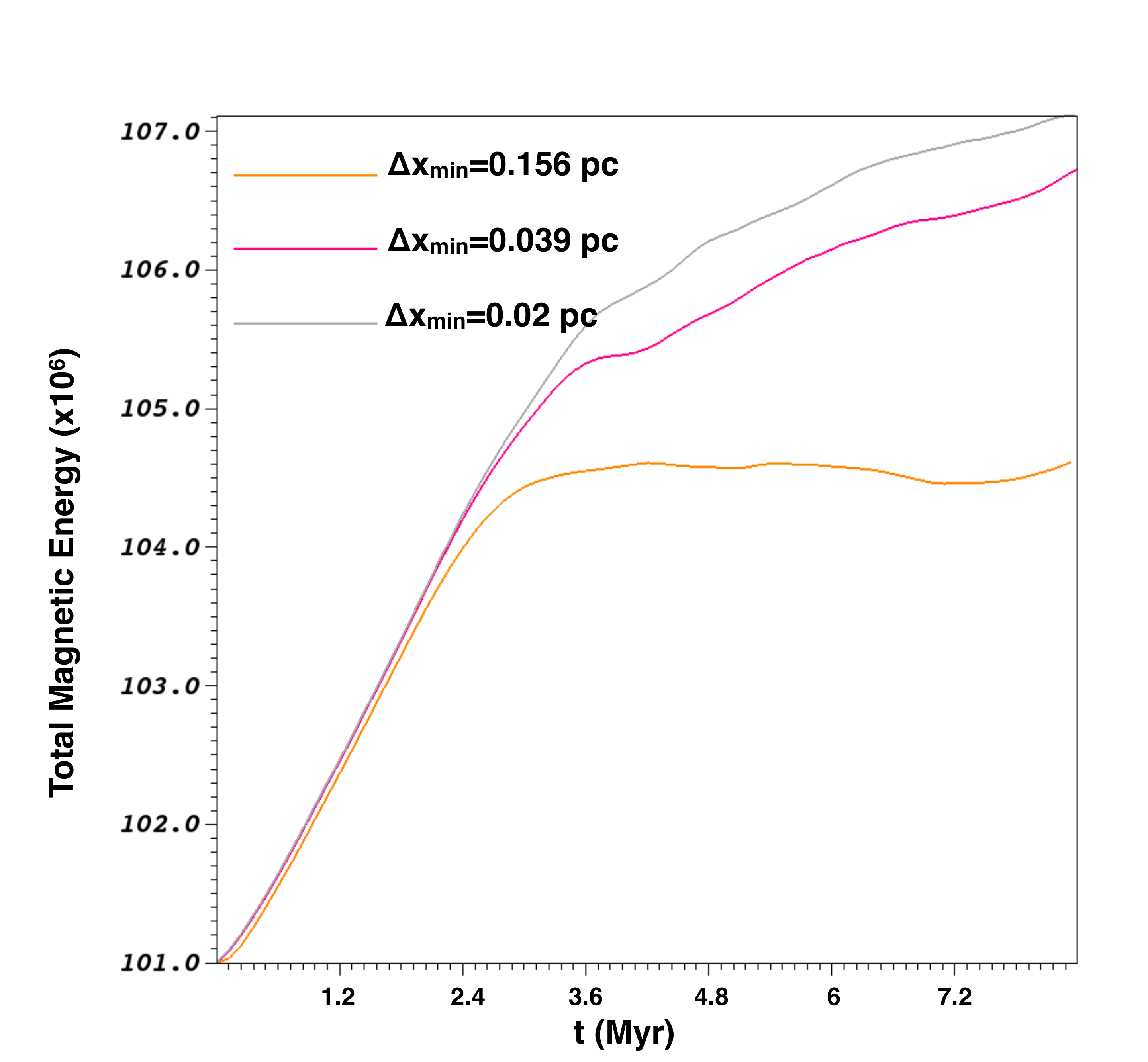}
    \caption {Total scaled magnetic energy over time for three different resolutions. Note that at higher resolutions, the amplification of the field by post-shock compression continues beyond $t>2.4~Myr$, where it tapers off at the resolution of the present paper.}
    \label{totalmagneticenergy}
\end{figure}

Second, the small-scale structure that develops in the post-shock flow has not converged at the resolution of the present paper. These results are consistent with \cite{koyama2004}, who show that thermal instability-induced fragmentation converges (in 1D) only when 1) the Field Length, $\lambda_F=(\kappa T/\rho^2\Lambda)^{1/2}$, where $\kappa$ is the coefficient of thermal conductivity and $\Lambda$ is the cooling rate of thermally unstable gas \citep{field1965}, is resolved by at least 3 zones, and 2) thermal conduction is included in the simulations. Along the collision interface, and away from the flow boundaries, the present suite of finite simulations are effectively 1D (at early times). However, since thermal conduction is ignored in the present work, we expect to see artificial growth of small-scale modes of the TI, seeded by numerical noise on the grid scale. This explains the formation of increasingly smaller-scale ‘clumplets’ in the post-shock flow with increasing resolution.

Taken together, a direct comparison of the simulation results to filaments in the ISM is not possible. However, our intention in this paper was to shed light on a potential MHD shock mechanism associated with oblique colliding flows, rather than investigate the detailed characteristics of filaments formed by this mechanism. Moreover, since the present suite of runs are 2D, the detailed structure of the filaments are artificially idealized from the outset (clearly evident when comparing the rightmost panel in Fig. \ref{convergencestudy} with its 3D counterpart: Fig. \ref{3Drun}). The results of this resolution study support that the reorientation mechanism is a large-scale phenomenon (which we argue is driven by post-shock pressure gradients), and that it is sufficiently resolved in the present paper.

\section*{Acknowledgments}

This research was supported in part by the US Department of Energy through grant GR523126, the National Science Foundation through grant GR506177, and the Space Telescope Science Institute through grant GR528562. A.P. would also like to acknowledge that partial salary support was provided by the Canadian Institute for Theoretical Astrophysics (CITA) National Fellowship.

\bibliography{erica}

\end{document}